\documentclass{article}

\makeatletter
\@ifundefined{lhs2tex.lhs2tex.sty.read}%
  {\@namedef{lhs2tex.lhs2tex.sty.read}{}%
   \newcommand\SkipToFmtEnd{}%
   \newcommand\EndFmtInput{}%
   \long\def\SkipToFmtEnd#1\EndFmtInput{}%
  }\SkipToFmtEnd

\newcommand\ReadOnlyOnce[1]{\@ifundefined{#1}{\@namedef{#1}{}}\SkipToFmtEnd}
\usepackage{amstext}
\usepackage{amssymb}
\usepackage{stmaryrd}
\DeclareFontFamily{OT1}{cmtex}{}
\DeclareFontShape{OT1}{cmtex}{m}{n}
  {<5><6><7><8>cmtex8
   <9>cmtex9
   <10><10.95><12><14.4><17.28><20.74><24.88>cmtex10}{}
\DeclareFontShape{OT1}{cmtex}{m}{it}
  {<-> ssub * cmtt/m/it}{}

\DeclareFontShape{OT1}{cmtt}{bx}{n}
  {<5><6><7><8>cmtt8
   <9>cmbtt9
   <10><10.95><12><14.4><17.28><20.74><24.88>cmbtt10}{}
\DeclareFontShape{OT1}{cmtex}{bx}{n}
  {<-> ssub * cmtt/bx/n}{}

\newcommand{\Conid}[1]{\mathit{#1}}
\newcommand{\Varid}[1]{\mathit{#1}}
\newcommand{\anonymous}{\kern0.06em \vbox{\hrule\@width.5em}}

\usepackage{polytable}

\@ifundefined{mathindent}%
  {\newdimen\mathindent\mathindent\leftmargini}%
  {}%

\def\resethooks{%
  \global\let\SaveRestoreHook\empty
  \global\let\ColumnHook\empty}
\newcommand*{\savecolumns}[1][default]%
  {\g@addto@macro\SaveRestoreHook{\savecolumns[#1]}}
\newcommand*{\restorecolumns}[1][default]%
  {\g@addto@macro\SaveRestoreHook{\restorecolumns[#1]}}
\newcommand*{\aligncolumn}[2]%
  {\g@addto@macro\ColumnHook{\column{#1}{#2}}}

\resethooks

\newcommand{\onelinecommentchars}{\quad-{}- }
\newcommand{\commentbeginchars}{\enskip\{-}
\newcommand{\commentendchars}{-\}\enskip}

\newcommand{\visiblecomments}{%
  \let\onelinecomment=\onelinecommentchars
  \let\commentbegin=\commentbeginchars
  \let\commentend=\commentendchars}

\newcommand{\invisiblecomments}{%
  \let\onelinecomment=\empty
  \let\commentbegin=\empty
  \let\commentend=\empty}

\visiblecomments

\newlength{\blanklineskip}
\newcommand{\hsindent}[1]{\quad}
\let\hspre\empty
\let\hspost\empty

\EndFmtInput
\makeatother
\ReadOnlyOnce{forall.fmt}%
\makeatletter

\let\HaskellResetHook\empty
\newcommand*{\AtHaskellReset}[1]{%
  \g@addto@macro\HaskellResetHook{#1}}
\newcommand*{\HaskellReset}{\HaskellResetHook}

\newcommand\hsforall{\global\let\hsdot=\hsperiodonce}
\newcommand*\hsperiodonce[2]{#2\global\let\hsdot=\hscompose}
\newcommand*\hscompose[2]{#1}

\AtHaskellReset{\global\let\hsdot=\hscompose}

\HaskellReset

\makeatother
\EndFmtInput
\ReadOnlyOnce{polycode.fmt}%
\makeatletter

\newcommand{\hsnewpar}[1]%
  {{\parskip=0pt\parindent=0pt\par\vskip #1\noindent}}

\newcommand{\hscodestyle}{}

\newcommand{\sethscode}[1]%
  {\expandafter\let\expandafter\hscode\csname #1\endcsname
   \expandafter\let\expandafter\endhscode\csname end#1\endcsname}

  {\par\noindent
   \advance\leftskip\mathindent
   \hscodestyle
   \let\\=\@normalcr
   \let\hspre\(\let\hspost\)%
   \pboxed}%
  {\endpboxed\)%
   \par\noindent
   \ignorespacesafterend}

  {\hsnewpar\abovedisplayskip
   \advance\leftskip\mathindent
   \hscodestyle
   \let\hspre\(\let\hspost\)%
   \pboxed}%
  {\endpboxed%
   \hsnewpar\belowdisplayskip
   \ignorespacesafterend}

  {\hsnewpar\abovedisplayskip
   \advance\leftskip\mathindent
   \hscodestyle
   \let\\=\@normalcr
   \(\pboxed}%
  {\endpboxed\)%
   \hsnewpar\belowdisplayskip
   \ignorespacesafterend}

\newcommand{\plainhs}{\sethscode{plainhscode}}

\plainhs

  {\hsnewpar\abovedisplayskip
   \advance\leftskip\mathindent
   \hscodestyle
   \let\\=\@normalcr
   \(\parray}%
  {\endparray\)%
   \hsnewpar\belowdisplayskip
   \ignorespacesafterend}

  {\parray}{\endparray}

  {\(\parray}{\endparray\)}

\def\codeframewidth{\arrayrulewidth}
\RequirePackage{calc}

  {\parskip=\abovedisplayskip\par\noindent
   \hscodestyle
   \arrayrulewidth=\codeframewidth
   \tabular{@{}|p{\linewidth-2\arraycolsep-2\arrayrulewidth-2pt}|@{}}%
   \hline\framedhslinecorrect\\{-1.5ex}%
   \let\endoflinesave=\\
   \let\\=\@normalcr
   \(\pboxed}%
  {\endpboxed\)%
   \framedhslinecorrect\endoflinesave{.5ex}\hline
   \endtabular
   \parskip=\belowdisplayskip\par\noindent
   \ignorespacesafterend}

\newcommand{\framedhslinecorrect}[2]%
  {#1[#2]}

  {\(\def\column##1##2{}%
   \let\>\undefined\let\<\undefined\let\\\undefined
   \newcommand\>[1][]{}\newcommand\<[1][]{}\newcommand\\[1][]{}%
   \def\fromto##1##2##3{##3}%
   }{\) }%

  {\let\orighscode=\hscode
   \let\origendhscode=\endhscode
   \def\endhscode{\def\hscode{\endgroup\def\@currenvir{hscode}\\}\begingroup}
   \orighscode\def\hscode{\endgroup\def\@currenvir{hscode}}}%
  {\origendhscode
   \global\let\hscode=\orighscode
   \global\let\endhscode=\origendhscode}%

\makeatother
\EndFmtInput

\ReadOnlyOnce{Formatting.fmt}%
\makeatletter

\let\Varid\mathit
\let\Conid\mathsf

\def\commentbegin{\quad\begingroup\color{Green}\{\ }
\def\commentend{\}\endgroup}

\makeatother
\EndFmtInput

\usepackage{classicthesis}
\usepackage{amsmath}
\usepackage{scalerel}
\usepackage{hyperref}
\usepackage{url}
\usepackage[dvipsnames]{xcolor}
\usepackage{subcaption}
\usepackage{graphicx}
\usepackage{natbib}
\usepackage[T1]{fontenc}
\usepackage{paralist}

\usepackage{doubleequals}

\newcommand{\delete}[1]{}

\allowdisplaybreaks

\newcommand{\txbr}[1]{\textcolor{Orange}{#1}}
\newcommand{\txtl}[1]{\textcolor{teal}{#1}}
\newcommand{\txbl}[1]{\textcolor{Cerulean}{#1}}

\newtheorem{theorem}{Theorem}

\newtheorem{lemma}[theorem]{Lemma}

\definecolor{mediumpersianblue}{rgb}{0.0, 0.4, 0.65}
\everymath{\color{mediumpersianblue}}

\begin{document}

\title{{\large\bf Functional Pearl}\\
Longest Segment of Balanced Parentheses:\\
{\large an Exercise in Program Inversion in a Segment Problem}}

\author{\color{black} Shin-Cheng Mu,~Tsung-Ju Chiang\\
    {\small\color{black}  Institute of Information Science, 
    Academia Sinica, Taiwan} 
    }
\date{}
\maketitle

\begin{abstract}
Given a string of parentheses, the task is to find the longest consecutive segment that is balanced, in linear time.
We find this problem interesting because it involves a combination of techniques: the usual approach for solving segment problems, and a theorem for constructing the inverse of a function --- through which we derive an instance of shift-reduce parsing.
\end{abstract}


\section{Introduction}
\label{sec:intro}

Given a string of parentheses, the task is to find a longest consecutive segment that is balanced.
For example, for input \ensuremath{\text{\ttfamily \char34 ))(()())())()(\char34}} the output should be \ensuremath{\text{\ttfamily \char34 (()())()\char34}}.
We also consider a reduced version of the problem in which we return only the length of the segment.
While there is no direct application of this problem
\footnote{However, the length-only version was possibly used as an interview problem collected in, for example, \url{https://leetcode.com/problems/longest-valid-parentheses/}.},
the authors find it interesting because it involves two techniques.
Firstly, derivation for such \emph{optimal segment} problems (those whose goal is to compute a segment of a list that is optimal up to certain criteria) usually follows a certain pattern (e.g. \cite{Bird:87:Introduction, Gibbons:97:Calculating, Zantema:92:Longest}). We would like to see how well that works for this case. Secondly, at one point we will need to construct the right inverse of a function. It will turn out that we will discover an instance of shift-reduce parsing.

\paragraph*{Specification}~ Balanced parentheses can be captured by a number of grammars, for example \ensuremath{\Conid{S}\to \epsilon\mid (\Conid{S})\mid \Conid{SS}}, or \ensuremath{\Conid{S}\to \Conid{T}^{*}} and \ensuremath{\Conid{T}\to (\Conid{S})}.
After trying some of them, the authors decided on
\begin{hscode}\SaveRestoreHook
\column{B}{@{}>{\hspre}l<{\hspost}@{}}%
\column{E}{@{}>{\hspre}l<{\hspost}@{}}%
\>[B]{}\Conid{S}\to \epsilon\mid (\Conid{S})\;\Conid{S}~~,{}\<[E]%
\ColumnHook
\end{hscode}\resethooks
because it is unambiguous and the most concise.
Other grammars have worked too, albeit leading to lengthier algorithms.
The parse tree of the chosen grammar can be represented in Haskell as below,
with a function \ensuremath{\Varid{pr}} specifying how a tree is printed:
\begin{hscode}\SaveRestoreHook
\column{B}{@{}>{\hspre}l<{\hspost}@{}}%
\column{15}{@{}>{\hspre}l<{\hspost}@{}}%
\column{E}{@{}>{\hspre}l<{\hspost}@{}}%
\>[B]{}\mathbf{data}\;\Conid{Tree}\mathrel{=}\Conid{Nul}\mid \Conid{Bin}\;\Conid{Tree}\;\Conid{Tree}~~,{}\<[E]%
\\[\blanklineskip]%
\>[B]{}\Varid{pr}\mathbin{::}\Conid{Tree}\to \Conid{String}{}\<[E]%
\\
\>[B]{}\Varid{pr}\;\Conid{Nul}{}\<[15]%
\>[15]{}\mathrel{=}\text{\ttfamily \char34 \char34}{}\<[E]%
\\
\>[B]{}\Varid{pr}\;(\Conid{Bin}\;\Varid{t}\;\Varid{u}){}\<[15]%
\>[15]{}\mathrel{=}\text{\ttfamily \char34 (\char34}\mathbin{+\!\!\!\!\!+}\Varid{pr}\;\Varid{t}\mathbin{+\!\!\!\!\!+}\text{\ttfamily \char34 )\char34}\mathbin{+\!\!\!\!\!+}\Varid{pr}\;\Varid{u}~~.{}\<[E]%
\ColumnHook
\end{hscode}\resethooks
For example, letting \ensuremath{\Varid{t}_{1}\mathrel{=}\Conid{Bin}\;\Conid{Nul}\;\Conid{Nul}} and \ensuremath{\Varid{t}_{2}\mathrel{=}\Conid{Bin}\;\Conid{Nul}\;(\Conid{Bin}\;\Conid{Nul}\;\Conid{Nul})},
we have
\ensuremath{\Varid{pr}\;\Varid{t}_{1}\mathrel{=}} {\tt "\txbr{()}"},
\ensuremath{\Varid{pr}\;\Varid{t}_{2}\mathrel{=}} {\tt "\txtl{()()}"}
and \ensuremath{\Varid{pr}\;(\Conid{Bin}\;\Varid{t}_{2}\;\Varid{t}_{1})\mathrel{=}} {\tt "(\txtl{()()})\txbr{()}"}
(parentheses are colored to aid the reader).

Function \ensuremath{\Varid{pr}} is injective but not surjective: it does not yield un-balanced strings.
Therefore its right inverse, that is, the function \ensuremath{{\Varid{pr}}^{\hstretch{0.5}{-}1}} such that \ensuremath{\Varid{pr}\;({\Varid{pr}}^{\hstretch{0.5}{-}1}\;\Varid{xs})\mathrel{=}\Varid{xs}}, is partial;
its domain is the set of balanced parenthesis strings.
We implement it by a function that is made total by using the \ensuremath{\Conid{Maybe}} monad.
This function \ensuremath{\Varid{parse}\mathbin{::}\Conid{String}\to \Conid{Maybe}\;\Conid{Tree}} builds a parse tree  --- \ensuremath{\Varid{parse}\;\Varid{xs}} should return \ensuremath{\Conid{Just}\;\Varid{t}} such that \ensuremath{\Varid{pr}\;\Varid{t}\mathrel{=}\Varid{xs}} if \ensuremath{\Varid{xs}} is balanced, and return \ensuremath{\Conid{Nothing}} otherwise.
While this defines \ensuremath{\Varid{parse}} already, a direct definition of \ensuremath{\Varid{parse}} will be presented in Section~\ref{sec:spine}.

The problem can then be specified as below, where \ensuremath{\Varid{lbs}} stands for ``longest balanced segment (of parentheses)'':
\begin{hscode}\SaveRestoreHook
\column{B}{@{}>{\hspre}l<{\hspost}@{}}%
\column{E}{@{}>{\hspre}l<{\hspost}@{}}%
\>[B]{}\Varid{lbs}\mathbin{::}\Conid{String}\to \Conid{Tree}{}\<[E]%
\\
\>[B]{}\Varid{lbs}\mathrel{=}\Varid{maxBy}\;\Varid{size}\mathbin{\cdot}\Varid{filtJust}\mathbin{\cdot}\Varid{map}\;\Varid{parse}\mathbin{\cdot}\Varid{segments}~~,{}\<[E]%
\\[\blanklineskip]%
\>[B]{}\Varid{segments}\mathrel{=}\Varid{concat}\mathbin{\cdot}\Varid{map}\;\Varid{inits}\mathbin{\cdot}\Varid{tails}~~,{}\<[E]%
\\
\>[B]{}\Varid{filtJust}\;\Varid{ts}\mathrel{=}[\mskip1.5mu \Varid{t}\mid \Conid{Just}\;\Varid{t}\leftarrow \Varid{ts}\mskip1.5mu]~~,{}\<[E]%
\\
\>[B]{}\Varid{size}\;\Varid{t}\mathrel{=}\Varid{length}\;(\Varid{pr}\;\Varid{t})~~.{}\<[E]%
\ColumnHook
\end{hscode}\resethooks
The function \ensuremath{\Varid{segments}\mathbin{::}[\mskip1.5mu \Varid{a}\mskip1.5mu]\to [\mskip1.5mu [\mskip1.5mu \Varid{a}\mskip1.5mu]\mskip1.5mu]} returns all segments of a list, with \ensuremath{\Varid{inits},\Varid{tails}\mathbin{::}[\mskip1.5mu \Varid{a}\mskip1.5mu]\to [\mskip1.5mu [\mskip1.5mu \Varid{a}\mskip1.5mu]\mskip1.5mu]} respectively computing all prefixes and suffixes of their input lists.
The result of \ensuremath{\Varid{map}\;\Varid{parse}} is passed to \ensuremath{\Varid{filtJust}\mathbin{::}[\mskip1.5mu \Conid{Maybe}\;\Varid{a}\mskip1.5mu]\to [\mskip1.5mu \Varid{a}\mskip1.5mu]}, which collects only those elements wrapped by \ensuremath{\Conid{Just}}.
For example, \ensuremath{\Varid{filtJust}\;[\mskip1.5mu \Conid{Just}\;\mathrm{1},\Conid{Nothing},\Conid{Just}\;\mathrm{2}\mskip1.5mu]\mathrel{=}[\mskip1.5mu \mathrm{1},\mathrm{2}\mskip1.5mu]}.
\footnote{\ensuremath{\Varid{filtJust}} is called \ensuremath{\Varid{catMaybes}} in the standard Haskell libraries. The authors think the name \ensuremath{\Varid{filtJust}} is more informative.}
For this problem \ensuremath{\Varid{filtJust}} always returns a non-empty list, because the empty string, which is a member of \ensuremath{\Varid{segments}\;\Varid{xs}} for any \ensuremath{\Varid{xs}}, can always be parsed to \ensuremath{\Conid{Just}\;\Conid{Nul}}.
Given \ensuremath{\Varid{f}\mathbin{::}\Varid{a}\to \Varid{b}} where \ensuremath{\Varid{b}} is a type that is ordered, \ensuremath{\Varid{maxBy}\;\Varid{f}\mathbin{::}[\mskip1.5mu \Varid{a}\mskip1.5mu]\to \Varid{a}} picks a maximum element from the input.
Finally, \ensuremath{\Varid{size}\;\Varid{t}} computes the length of \ensuremath{\Varid{pr}\;\Varid{t}}.

The length-only problem can be specified by \ensuremath{\Varid{lbsl}\mathrel{=}\Varid{size}\mathbin{\cdot}\Varid{lbs}}.

\section{The Prefix-Suffix Decomposition}
\label{sec:inits-tails}

It is known that many optimal segment problems can be solved by following a fixed pattern
\citep{Bird:87:Introduction, Gibbons:97:Calculating, Zantema:92:Longest},
which we refer to as \emph{prefix-suffix decomposition}.
In the first step, finding an optimal segment is factored into finding, for each suffix, an optimal prefix.
For our problem, the calculation goes:
\begin{hscode}\SaveRestoreHook
\column{B}{@{}>{\hspre}l<{\hspost}@{}}%
\column{7}{@{}>{\hspre}l<{\hspost}@{}}%
\column{9}{@{}>{\hspre}l<{\hspost}@{}}%
\column{E}{@{}>{\hspre}l<{\hspost}@{}}%
\>[7]{}\Varid{maxBy}\;\Varid{size}\mathbin{\cdot}\Varid{filtJust}\mathbin{\cdot}\Varid{map}\;\Varid{parse}\mathbin{\cdot}\Varid{segments}{}\<[E]%
\\
\>[B]{}\mathbin{=}{}\<[9]%
\>[9]{}\mbox{\commentbegin  definition of \ensuremath{\Varid{segments}}  \commentend}{}\<[E]%
\\
\>[B]{}\hsindent{7}{}\<[7]%
\>[7]{}\Varid{maxBy}\;\Varid{size}\mathbin{\cdot}\Varid{filtJust}\mathbin{\cdot}\Varid{map}\;\Varid{parse}\mathbin{\cdot}\Varid{concat}\mathbin{\cdot}\Varid{map}\;\Varid{inits}\mathbin{\cdot}\Varid{tails}{}\<[E]%
\\
\>[B]{}\mathbin{=}{}\<[9]%
\>[9]{}\mbox{\commentbegin  since \ensuremath{\Varid{map}\;\Varid{f}\mathbin{\cdot}\Varid{concat}\mathrel{=}\Varid{concat}\mathbin{\cdot}\Varid{map}\;(\Varid{map}\;\Varid{f})}, \ensuremath{\Varid{map}} fusion  \commentend}{}\<[E]%
\\
\>[B]{}\hsindent{7}{}\<[7]%
\>[7]{}\Varid{maxBy}\;\Varid{size}\mathbin{\cdot}\Varid{filtJust}\mathbin{\cdot}\Varid{concat}\mathbin{\cdot}\Varid{map}\;(\Varid{map}\;\Varid{parse}\mathbin{\cdot}\Varid{inits})\mathbin{\cdot}\Varid{tails}{}\<[E]%
\\
\>[B]{}\mathbin{=}{}\<[9]%
\>[9]{}\mbox{\commentbegin  since \ensuremath{\Varid{filtJust}\mathbin{\cdot}\Varid{concat}\mathrel{=}\Varid{concat}\mathbin{\cdot}\Varid{map}\;\Varid{filtJust}}  \commentend}{}\<[E]%
\\
\>[B]{}\hsindent{7}{}\<[7]%
\>[7]{}\Varid{maxBy}\;\Varid{size}\mathbin{\cdot}\Varid{concat}\mathbin{\cdot}\Varid{map}\;(\Varid{filtJust}\mathbin{\cdot}\Varid{map}\;\Varid{parse}\mathbin{\cdot}\Varid{inits})\mathbin{\cdot}\Varid{tails}{}\<[E]%
\\
\>[B]{}\mathbin{=}{}\<[9]%
\>[9]{}\mbox{\commentbegin  since \ensuremath{\Varid{maxBy}\;\Varid{f}\mathbin{\cdot}\Varid{concat}\mathrel{=}\Varid{maxBy}\;\Varid{f}\mathbin{\cdot}\Varid{map}\;(\Varid{maxBy}\;\Varid{f})}  \commentend}{}\<[E]%
\\
\>[B]{}\hsindent{7}{}\<[7]%
\>[7]{}\Varid{maxBy}\;\Varid{size}\mathbin{\cdot}\Varid{map}\;(\Varid{maxBy}\;\Varid{size}\mathbin{\cdot}\Varid{filtJust}\mathbin{\cdot}\Varid{map}\;\Varid{parse}\mathbin{\cdot}\Varid{inits})\mathbin{\cdot}\Varid{tails}~~.{}\<[E]%
\ColumnHook
\end{hscode}\resethooks
For each suffix returned by \ensuremath{\Varid{tails}}, the program above computes its longest \emph{prefix} of balanced parentheses by \ensuremath{\Varid{maxBy}\;\Varid{size}\mathbin{\cdot}\Varid{filtJust}\mathbin{\cdot}\Varid{map}\;\Varid{parse}\mathbin{\cdot}\Varid{inits}}. We abbreviate the latter to \ensuremath{\Varid{lbp}} (for ``longest balanced prefix'').

Generating every suffix and computing \ensuremath{\Varid{lbp}} for each of them is rather costly.
The next step is to try to apply the following \emph{scan lemma},
which says that if a function \ensuremath{\Varid{f}} can be expressed as right fold,
there is a more efficient algorithm to compute \ensuremath{\Varid{map}\;\Varid{f}\mathbin{\cdot}\Varid{tails}}:
\begin{lemma}
\label{lma:scan-lemma}
{\rm
\ensuremath{\Varid{map}\;(\Varid{foldr}\;(\oplus)\;\Varid{e})\mathbin{\cdot}\Varid{tails}\mathrel{=}\Varid{scanr}\;(\oplus)\;\Varid{e}}, where
\begin{hscode}\SaveRestoreHook
\column{B}{@{}>{\hspre}l<{\hspost}@{}}%
\column{24}{@{}>{\hspre}l<{\hspost}@{}}%
\column{E}{@{}>{\hspre}l<{\hspost}@{}}%
\>[B]{}\Varid{scanr}\mathbin{::}(\Varid{a}\to \Varid{b}\to \Varid{b})\to \Varid{b}\to [\mskip1.5mu \Varid{a}\mskip1.5mu]\to [\mskip1.5mu \Varid{b}\mskip1.5mu]{}\<[E]%
\\
\>[B]{}\Varid{scanr}\;(\oplus)\;\Varid{e}\;[\mskip1.5mu \mskip1.5mu]{}\<[24]%
\>[24]{}\mathrel{=}[\mskip1.5mu \Varid{e}\mskip1.5mu]{}\<[E]%
\\
\>[B]{}\Varid{scanr}\;(\oplus)\;\Varid{e}\;(\Varid{x}\mathbin{:}\Varid{xs}){}\<[24]%
\>[24]{}\mathrel{=}\mathbf{let}\;(\Varid{y}\mathbin{:}\Varid{ys})\mathrel{=}\Varid{scanr}\;(\oplus)\;\Varid{e}\;\Varid{xs}\;\mathbf{in}\;(\Varid{x}\mathbin{\oplus}\Varid{y})\mathbin{:}\Varid{y}\mathbin{:}\Varid{ys}~~.{}\<[E]%
\ColumnHook
\end{hscode}\resethooks
} 
\end{lemma}
If \ensuremath{\Varid{lbp}} can be written in the form \ensuremath{\Varid{foldr}\;(\oplus)\;\Varid{e}}, we do not need to compute \ensuremath{\Varid{lbp}} of each suffix from scratch;
each optimal prefix can be computed, in \ensuremath{\Varid{scanr}}, from the previous optimal prefix by \ensuremath{(\oplus)}.
If \ensuremath{(\oplus)} is a constant-time operation, we get a linear-time algorithm.

The next challenge is therefore to express \ensuremath{\Varid{lbp}} as a right fold.
Since \ensuremath{\Varid{inits}} can be expressed as a right fold  --- \ensuremath{\Varid{inits}\mathrel{=}\Varid{foldr}\;(\lambda \Varid{x}\;\Varid{xss}\to [\mskip1.5mu \mskip1.5mu]\mathbin{:}\Varid{map}\;(\Varid{x}\mathbin{:})\;\Varid{xss})\;[\mskip1.5mu [\mskip1.5mu \mskip1.5mu]\mskip1.5mu]},
a reasonable attempt is to fuse \ensuremath{\Varid{maxBy}\;\Varid{size}\mathbin{\cdot}\Varid{filtJust}\mathbin{\cdot}\Varid{map}\;\Varid{parse}} with \ensuremath{\Varid{inits}}, to form a single \ensuremath{\Varid{foldr}}.
Recall the \ensuremath{\Varid{foldr}}-fusion theorem:
\begin{theorem}[\ensuremath{\Varid{foldr}}-fusion]
\label{thm:foldr-fusion}
{\rm
  \ensuremath{\Varid{h}\mathbin{\cdot}\Varid{foldr}\;\Varid{f}\;\Varid{e}\mathrel{=}\Varid{foldr}\;\Varid{g}\;(\Varid{h}\;\Varid{e})} if \ensuremath{\Varid{h}\;(\Varid{f}\;\Varid{x}\;\Varid{y})\mathrel{=}\Varid{g}\;\Varid{x}\;(\Varid{h}\;\Varid{y})}.
} 
\end{theorem}
The antecedent \ensuremath{\Varid{h}\;(\Varid{f}\;\Varid{x}\;\Varid{y})\mathrel{=}\Varid{g}\;\Varid{x}\;(\Varid{h}\;\Varid{y})} will be referred to as the \emph{fusion condition}.
To fuse \ensuremath{\Varid{map}\;\Varid{parse}} and \ensuremath{\Varid{inits}} using Theorem~\ref{thm:foldr-fusion},
we calculate from the LHS of the fusion condition (with \ensuremath{\Varid{h}\mathrel{=}\Varid{map}\;\Varid{parse}} and \ensuremath{\Varid{f}\mathrel{=}(\lambda \Varid{x}\;\Varid{xss}\to [\mskip1.5mu \mskip1.5mu]\mathbin{:}\Varid{map}\;(\Varid{x}\mathbin{:})\;\Varid{xss})}):
\begin{hscode}\SaveRestoreHook
\column{B}{@{}>{\hspre}l<{\hspost}@{}}%
\column{7}{@{}>{\hspre}l<{\hspost}@{}}%
\column{9}{@{}>{\hspre}l<{\hspost}@{}}%
\column{28}{@{}>{\hspre}l<{\hspost}@{}}%
\column{E}{@{}>{\hspre}l<{\hspost}@{}}%
\>[7]{}\Varid{map}\;\Varid{parse}\;([\mskip1.5mu \mskip1.5mu]\mathbin{:}\Varid{map}\;(\Varid{x}\mathbin{:})\;\Varid{xss}){}\<[E]%
\\
\>[B]{}\mathbin{=}{}\<[9]%
\>[9]{}\mbox{\commentbegin  since \ensuremath{\Varid{parse}\;[\mskip1.5mu \mskip1.5mu]\mathrel{=}\Conid{Just}\;\Conid{Nul}}  \commentend}{}\<[E]%
\\
\>[B]{}\hsindent{7}{}\<[7]%
\>[7]{}\Conid{Just}\;\Conid{Nul}\mathbin{:}\Varid{map}\;(\Varid{parse}\mathbin{\cdot}(\Varid{x}\mathbin{:}))\;\Varid{xss}{}\<[E]%
\\
\>[B]{}\mathbin{=}{}\<[9]%
\>[9]{}\mbox{\commentbegin  wish, for some \ensuremath{\Varid{g'}}  \commentend}{}\<[E]%
\\
\>[B]{}\hsindent{7}{}\<[7]%
\>[7]{}\Conid{Just}\;\Conid{Nul}\mathbin{:}\Varid{g'}\;\Varid{x}\;(\Varid{map}\;\Varid{parse}\;\Varid{xss}){}\<[E]%
\\
\>[B]{}\mathbin{=}{}\<[9]%
\>[9]{}\mbox{\commentbegin  let \ensuremath{\Varid{g}\;\Varid{x}\;\Varid{ts}\mathrel{=}\Conid{Just}\;\Conid{Nul}\mathbin{:}\Varid{g}\;\Varid{x}\;\Varid{ts}}  \commentend}{}\<[E]%
\\
\>[B]{}\hsindent{7}{}\<[7]%
\>[7]{}\Varid{g}\;\Varid{x}\;(\Varid{map}\;\Varid{parse}\;\Varid{xss}){}\<[28]%
\>[28]{}~~.{}\<[E]%
\ColumnHook
\end{hscode}\resethooks
We can construct \ensuremath{\Varid{g}} if (and only if) there is a function \ensuremath{\Varid{g'}} such that \ensuremath{\Varid{g'}\;\Varid{x}\;(\Varid{map}\;\Varid{parse}\;\Varid{xss})\mathrel{=}\Varid{map}\;(\Varid{parse}\mathbin{\cdot}(\Varid{x}\mathbin{:}))\;\Varid{xss}}.
Is that possible?

It is not hard to see that the answer is no. Consider \ensuremath{\Varid{xss}\mathrel{=}[\mskip1.5mu \text{\ttfamily \char34 )\char34},\text{\ttfamily \char34 )()\char34}\mskip1.5mu]} and \ensuremath{\Varid{x}\mathrel{=}\text{\ttfamily '('}}. Since both strings in \ensuremath{\Varid{xss}} are not balanced, \ensuremath{\Varid{map}\;\Varid{parse}\;\Varid{xss}} gives us \ensuremath{[\mskip1.5mu \Conid{Nothing},\Conid{Nothing}\mskip1.5mu]}. However, \ensuremath{\Varid{map}\;(\Varid{x}\mathbin{:})\;\Varid{xss}\mathrel{=}[\mskip1.5mu \text{\ttfamily \char34 ()\char34},\text{\ttfamily \char34 ()()\char34}\mskip1.5mu]}, a list of balanced strings. Therefore \ensuremath{\Varid{g'}} has to produce something from nothing --- an impossible task. We have to generalise our problem such that \ensuremath{\Varid{g'}} receives inputs that are more informative.




\section{Partially Balanced Strings}
\label{sec:partial-balance}

A string of parentheses is said to be \emph{left-partially balanced} if it may possibly be balanced by adding zero or more parentheses to the left.
For example, \ensuremath{\Varid{xs}\mathrel{=}}{\tt "\txtl{(())()}\txbr{))}\txbl{()}"} is left-partially balanced because \ensuremath{\Varid{txbr}\;\text{\ttfamily \char34 ((\char34}\mathbin{+\!\!\!\!\!+}\Varid{xs}} is balanced --- again we use coloring to help the reader parsing the string.
Note that {\tt "\txbr{(}()\txbr{(}"} \ensuremath{\mathbin{+\!\!\!\!\!+}} \ensuremath{\Varid{xs}} is also balanced.
For a counter example, the string \ensuremath{\Varid{ys}\mathrel{=}} {\tt "()\txbr{(}()"} is not left-partially balanced --- due to the unmatched {\tt '\txbr{(}'} in the middle of \ensuremath{\Varid{ys}}, there is no \ensuremath{\Varid{zs}} such that \ensuremath{\Varid{zs}\mathbin{+\!\!\!\!\!+}\Varid{ys}} can be balanced.

While parsing a \emph{fully balanced} string cannot be expressed as a right fold, it is possible to parse \emph{left-partially balanced} strings using a right fold.
In this section we consider what data structure such a string should be parsed to. We discuss how to how to parse it in the next section.

A left-partially balanced string can always be uniquely factored into a sequence of fully balanced substrings, separated by one or more right parentheses.
For example, \ensuremath{\Varid{xs}} can be factored into two balanced substrings, {\tt "\txtl{(())()}"} and {\tt"\txbl{()}"}, separated by {\tt"\txbr{))}"}.
One of the possible ways to represent such a string is by a list of trees --- a \ensuremath{\Conid{Forest}}, where the trees are supposed to be separated by a \ensuremath{\text{\ttfamily ')'}}.
That is, such a forest can be printed by:
\newcommand{\hscomment}[1]{\{-~~\mbox{#1}~~-\}}
\begin{hscode}\SaveRestoreHook
\column{B}{@{}>{\hspre}l<{\hspost}@{}}%
\column{13}{@{}>{\hspre}l<{\hspost}@{}}%
\column{E}{@{}>{\hspre}l<{\hspost}@{}}%
\>[B]{}\mathbf{type}\;\Conid{Forest}\mathrel{=}[\mskip1.5mu \Conid{Tree}\mskip1.5mu]~~,\qquad\qquad \hscomment{non-empty}{}\<[E]%
\\[\blanklineskip]%
\>[B]{}\Varid{prF}\mathbin{::}\Conid{Forest}\to \Conid{String}{}\<[E]%
\\
\>[B]{}\Varid{prF}\;[\mskip1.5mu \Varid{t}\mskip1.5mu]{}\<[13]%
\>[13]{}\mathrel{=}\Varid{pr}\;\Varid{t}{}\<[E]%
\\
\>[B]{}\Varid{prF}\;(\Varid{t}\mathbin{:}\Varid{ts}){}\<[13]%
\>[13]{}\mathrel{=}\Varid{pr}\;\Varid{t}\mathbin{+\!\!\!\!\!+}\text{\ttfamily \char34 )\char34}\mathbin{+\!\!\!\!\!+}\Varid{prF}\;\Varid{ts}~~.{}\<[E]%
\ColumnHook
\end{hscode}\resethooks
For example, \ensuremath{\Varid{xs}\mathrel{=}} {\tt "\txtl{(())()}\txbr{))}\txbl{()}"} can be represented by a forest containing three trees:
\begin{hscode}\SaveRestoreHook
\column{B}{@{}>{\hspre}l<{\hspost}@{}}%
\column{E}{@{}>{\hspre}l<{\hspost}@{}}%
\>[B]{}\Varid{ts}\mathrel{=}[\mskip1.5mu \Varid{txtl}\;(\Conid{Bin}\;(\Conid{Bin}\;\Conid{Nul}\;\Conid{Nul})\;(\Conid{Bin}\;\Conid{Nul}\;\Conid{Nul})),~~\Conid{Nul},~~\Varid{txbl}\;(\Conid{Bin}\;\Conid{Nul}\;\Conid{Nul})\mskip1.5mu]~~,{}\<[E]%
\ColumnHook
\end{hscode}\resethooks
where \ensuremath{\Varid{txtl}\;(\Conid{Bin}\;(\Conid{Bin}\;\Conid{Nul}\;\Conid{Nul})\;(\Conid{Bin}\;\Conid{Nul}\;\Conid{Nul}))} prints to {\tt "\txtl{(())()}"},
\ensuremath{\Varid{txbl}\;(\Conid{Bin}\;\Conid{Nul}\;\Conid{Nul})} prints to {\tt"\txbl{()}"}, and there is a \ensuremath{\Conid{Nul}} between
them due to the consecutive right parentheses {\tt"\txbr{))}"} in \ensuremath{\Varid{xs}}
(\ensuremath{\Conid{Nul}} itself prints to \ensuremath{\text{\ttfamily \char34 \char34}}).
One can verify that \ensuremath{\Varid{prF}\;\Varid{ts}\mathrel{=}\Varid{xs}} indeed.
Note that we let the type \ensuremath{\Conid{Forest}} be \emph{non-empty} lists of trees.
\footnote{We can let the non-emptiness be more explicit by letting \ensuremath{\Conid{Forest}\mathrel{=}(\Conid{Tree},[\mskip1.5mu \Conid{Tree}\mskip1.5mu])}. Presentation-wise, both representations have their pros and cons, and we eventually decided on using a list.}
The empty string can be represented by \ensuremath{[\mskip1.5mu \Conid{Nul}\mskip1.5mu]}, since \ensuremath{\Varid{prF}\;[\mskip1.5mu \Conid{Nul}\mskip1.5mu]\mathrel{=}\Varid{pr}\;\Conid{Nul}\mathrel{=}\text{\ttfamily \char34 \char34}}.

The aim now is to construct the right inverse of \ensuremath{\Varid{prF}},
such that a left-partially balanced string can be parsed using a right fold.

%


\section{Parsing Partially Balanced Strings}
\label{sec:spine}

Given a function \ensuremath{\Varid{f}\mathbin{::}\Varid{b}\to \Varid{t}}, the converse-of-a-function theorem \citep{BirddeMoor:97:Algebra, deMoorGibbons:00:Pointwise} constructs the relational converse --- a generalised notion of inverse --- of \ensuremath{\Varid{f}}.
The converse is given as a relational fold whose input type is \ensuremath{\Varid{t}}, which can be any inductively-defined datatype with a polynomial base functor.
We specialise the general theorem to our needs: we use it to construct only functions, not relations, and only for the case where \ensuremath{\Varid{t}} is a list type.
\begin{theorem}
\label{thm:conv-fun}
{\rm
Given \ensuremath{\Varid{f}\mathbin{::}\Varid{b}\to [\mskip1.5mu \Varid{a}\mskip1.5mu]}, if we have \ensuremath{\Varid{base}\mathbin{::}\Varid{b}} and \ensuremath{\Varid{step}\mathbin{::}\Varid{a}\to \Varid{b}\to \Varid{b}} satisfying:
\begin{align*}
\ensuremath{\Varid{f}\;\Varid{base}} &= \ensuremath{[\mskip1.5mu \mskip1.5mu]} \quad\wedge \\
\ensuremath{\Varid{f}\;(\Varid{step}\;\Varid{x}\;\Varid{t})} &= \ensuremath{\Varid{x}\mathbin{:}\Varid{f}\;\Varid{t}} \mbox{~~,}
\end{align*}
then \ensuremath{{\Varid{f}}^{\hstretch{0.5}{-}1}\mathrel{=}\Varid{foldr}\;\Varid{step}\;\Varid{base}} is a partial right inverse of \ensuremath{\Varid{f}}. That is, we have \ensuremath{\Varid{f}\;({\Varid{f}}^{\hstretch{0.5}{-}1}\;\Varid{xs})\mathrel{=}\Varid{xs}} for all \ensuremath{\Varid{xs}} in the range of \ensuremath{\Varid{f}}.
} 
\end{theorem}

While the general version of the theorem is not trivial to prove,
the version above, specialised to functions and lists, can be verified by an easy induction on the input list.

Recall that we wish to construct the right inverse of \ensuremath{\Varid{prF}} using Theorem~\ref{thm:conv-fun}.
It will be easier if we first construct a new definition of \ensuremath{\Varid{prF}}, one that is inductive, does not use \ensuremath{(\mathbin{+\!\!\!\!\!+})}, and does not rely on \ensuremath{\Varid{pr}}.
For a base case, \ensuremath{\Varid{prF}\;[\mskip1.5mu \Conid{Nul}\mskip1.5mu]\mathrel{=}\text{\ttfamily \char34 \char34}}.
It is also immediate that \ensuremath{\Varid{prF}\;(\Conid{Nul}\mathbin{:}\Varid{ts})\mathrel{=}\text{\ttfamily ')'}\mathbin{:}\Varid{prF}\;\Varid{ts}}.
When the list contains more than one tree and the first tree is not \ensuremath{\Conid{Nul}}, we calculate:
\begin{hscode}\SaveRestoreHook
\column{B}{@{}>{\hspre}l<{\hspost}@{}}%
\column{7}{@{}>{\hspre}l<{\hspost}@{}}%
\column{9}{@{}>{\hspre}l<{\hspost}@{}}%
\column{E}{@{}>{\hspre}l<{\hspost}@{}}%
\>[7]{}\Varid{prF}\;(\Conid{Bin}\;\Varid{t}\;\Varid{u}\mathbin{:}\Varid{ts}){}\<[E]%
\\
\>[B]{}\mathbin{=}{}\<[9]%
\>[9]{}\mbox{\commentbegin  definitions of \ensuremath{\Varid{pr}} and \ensuremath{\Varid{prF}}  \commentend}{}\<[E]%
\\
\>[B]{}\hsindent{7}{}\<[7]%
\>[7]{}\text{\ttfamily \char34 (\char34}\mathbin{+\!\!\!\!\!+}\Varid{pr}\;\Varid{t}\mathbin{+\!\!\!\!\!+}\text{\ttfamily \char34 )\char34}\mathbin{+\!\!\!\!\!+}\Varid{pr}\;\Varid{u}\mathbin{+\!\!\!\!\!+}\text{\ttfamily \char34 )\char34}\mathbin{+\!\!\!\!\!+}\Varid{prF}\;\Varid{ts}{}\<[E]%
\\
\>[B]{}\mathbin{=}{}\<[9]%
\>[9]{}\mbox{\commentbegin  definition of \ensuremath{\Varid{prF}}  \commentend}{}\<[E]%
\\
\>[B]{}\hsindent{7}{}\<[7]%
\>[7]{}\text{\ttfamily '('}\mathbin{:}\Varid{prF}\;(\Varid{t}\mathbin{:}\Varid{u}\mathbin{:}\Varid{ts})~~.{}\<[E]%
\ColumnHook
\end{hscode}\resethooks
We have thus derived the following new definition of \ensuremath{\Varid{prF}}:
\begin{hscode}\SaveRestoreHook
\column{B}{@{}>{\hspre}l<{\hspost}@{}}%
\column{20}{@{}>{\hspre}l<{\hspost}@{}}%
\column{E}{@{}>{\hspre}l<{\hspost}@{}}%
\>[B]{}\Varid{prF}\;[\mskip1.5mu \Conid{Nul}\mskip1.5mu]{}\<[20]%
\>[20]{}\mathrel{=}\text{\ttfamily \char34 \char34}{}\<[E]%
\\
\>[B]{}\Varid{prF}\;(\Conid{Nul}\mathbin{:}\Varid{ts}){}\<[20]%
\>[20]{}\mathrel{=}\text{\ttfamily ')'}\mathbin{:}\Varid{prF}\;\Varid{ts}{}\<[E]%
\\
\>[B]{}\Varid{prF}\;(\Conid{Bin}\;\Varid{t}\;\Varid{u}\mathbin{:}\Varid{ts}){}\<[20]%
\>[20]{}\mathrel{=}\text{\ttfamily '('}\mathbin{:}\Varid{prF}\;(\Varid{t}\mathbin{:}\Varid{u}\mathbin{:}\Varid{ts})~~.{}\<[E]%
\ColumnHook
\end{hscode}\resethooks

We are now ready to invert \ensuremath{\Varid{prF}} by Theorem~\ref{thm:conv-fun},
which amounts to finding \ensuremath{\Varid{base}} and \ensuremath{\Varid{step}} such that \ensuremath{\Varid{prF}\;\Varid{base}\mathrel{=}\text{\ttfamily \char34 \char34}} and \ensuremath{\Varid{prF}\;(\Varid{step}\;\Varid{x}\;\Varid{ts})\mathrel{=}\Varid{x}\mathbin{:}\Varid{prF}\;\Varid{ts}} for \ensuremath{\Varid{x}\mathrel{=}\text{\ttfamily '('}} or \ensuremath{\text{\ttfamily ')'}}.
With the inductive definition of \ensuremath{\Varid{prF}} in mind, we pick \ensuremath{\Varid{base}\mathrel{=}[\mskip1.5mu \Conid{Nul}\mskip1.5mu]}, and the following \ensuremath{\Varid{step}} meets the requirement:
\begin{hscode}\SaveRestoreHook
\column{B}{@{}>{\hspre}l<{\hspost}@{}}%
\column{24}{@{}>{\hspre}c<{\hspost}@{}}%
\column{24E}{@{}l@{}}%
\column{27}{@{}>{\hspre}l<{\hspost}@{}}%
\column{E}{@{}>{\hspre}l<{\hspost}@{}}%
\>[B]{}\Varid{step}\;\text{\ttfamily ')'}\;\Varid{ts}{}\<[24]%
\>[24]{}\mathrel{=}{}\<[24E]%
\>[27]{}\Conid{Nul}\mathbin{:}\Varid{ts}{}\<[E]%
\\
\>[B]{}\Varid{step}\;\text{\ttfamily '('}\;(\Varid{t}\mathbin{:}\Varid{u}\mathbin{:}\Varid{ts}){}\<[24]%
\>[24]{}\mathrel{=}{}\<[24E]%
\>[27]{}\Conid{Bin}\;\Varid{t}\;\Varid{u}\mathbin{:}\Varid{ts}~~.{}\<[E]%
\ColumnHook
\end{hscode}\resethooks
We have thus constructed \ensuremath{{\Varid{prF}}^{\hstretch{0.5}{-}1}\mathrel{=}\Varid{foldr}\;\Varid{step}\;[\mskip1.5mu \Conid{Nul}\mskip1.5mu]}.
If we expand the definitions, we have
\begin{hscode}\SaveRestoreHook
\column{B}{@{}>{\hspre}l<{\hspost}@{}}%
\column{16}{@{}>{\hspre}l<{\hspost}@{}}%
\column{E}{@{}>{\hspre}l<{\hspost}@{}}%
\>[B]{}{\Varid{prF}}^{\hstretch{0.5}{-}1}\mathbin{::}\Conid{String}\to \Conid{Forest}{}\<[E]%
\\
\>[B]{}{\Varid{prF}}^{\hstretch{0.5}{-}1}\;\text{\ttfamily \char34 \char34}{}\<[16]%
\>[16]{}\mathrel{=}[\mskip1.5mu \Conid{Nul}\mskip1.5mu]{}\<[E]%
\\
\>[B]{}{\Varid{prF}}^{\hstretch{0.5}{-}1}\;(\text{\ttfamily ')'}\mathbin{:}\Varid{xs}){}\<[16]%
\>[16]{}\mathrel{=}\Conid{Nul}\mathbin{:}{\Varid{prF}}^{\hstretch{0.5}{-}1}\;\Varid{xs}{}\<[E]%
\\
\>[B]{}{\Varid{prF}}^{\hstretch{0.5}{-}1}\;(\text{\ttfamily '('}\mathbin{:}\Varid{xs}){}\<[16]%
\>[16]{}\mathrel{=}\mathbf{case}\;{\Varid{prF}}^{\hstretch{0.5}{-}1}\;\Varid{xs}\;\mathbf{of}\;(\Varid{t}\mathbin{:}\Varid{u}\mathbin{:}\Varid{ts})\to \Conid{Bin}\;\Varid{t}\;\Varid{u}\mathbin{:}\Varid{ts}~~,{}\<[E]%
\ColumnHook
\end{hscode}\resethooks
which is pleasingly symmetrical to the inductive definition of \ensuremath{\Varid{prF}}.

For an operational explanation,
a right parenthesis \ensuremath{\text{\ttfamily ')'}} indicates starting a new tree, thus we start freshly with a \ensuremath{\Conid{Nul}};
a left parenthesis \ensuremath{\text{\ttfamily '('}} ought to be the leftmost symbol of some \ensuremath{\text{\ttfamily \char34 (t)u\char34}},
thus we wrap the two most recent siblings into one tree.
When there are no such two siblings (that is, \ensuremath{{\Varid{prF}}^{\hstretch{0.5}{-}1}\;\Varid{xs}} in the \ensuremath{\mathbf{case}} expression evaluates to a singleton list), the input \ensuremath{\text{\ttfamily '('}\mathbin{:}\Varid{xs}} is not a left-partially balanced string --- \ensuremath{\text{\ttfamily '('}} appears too early, and
the result is undefined.

Readers may have noticed the similarity to shift-reduce parsing,
in which, after reading a symbol we either "shift"
the symbol by pushing it onto a stack, or "reduce" the symbol against
a top segment of the stack.
Here, the forest is the stack.
The input is processed right-to-left, as opposed to left-to-right, which is more common when talking about parsing.
We shall discuss this issue further in Section~\ref{sec:conclude}.


We could proceed to work with \ensuremath{{\Varid{prF}}^{\hstretch{0.5}{-}1}} for the rest of this pearl but,
for clarity, we prefer to make the partiality explicit.
Let \ensuremath{\Varid{parseF}} be the monadified version of \ensuremath{{\Varid{prF}}^{\hstretch{0.5}{-}1}}, given by:
\begin{hscode}\SaveRestoreHook
\column{B}{@{}>{\hspre}l<{\hspost}@{}}%
\column{3}{@{}>{\hspre}l<{\hspost}@{}}%
\column{10}{@{}>{\hspre}l<{\hspost}@{}}%
\column{16}{@{}>{\hspre}l<{\hspost}@{}}%
\column{34}{@{}>{\hspre}l<{\hspost}@{}}%
\column{E}{@{}>{\hspre}l<{\hspost}@{}}%
\>[B]{}\Varid{parseF}\mathbin{::}\Conid{String}\to \Conid{Maybe}\;\Conid{Forest}{}\<[E]%
\\
\>[B]{}\Varid{parseF}\;\text{\ttfamily \char34 \char34}{}\<[16]%
\>[16]{}\mathrel{=}\Conid{Just}\;[\mskip1.5mu \Conid{Nul}\mskip1.5mu]{}\<[E]%
\\
\>[B]{}\Varid{parseF}\;(\Varid{x}\mathbin{:}\Varid{xs}){}\<[16]%
\>[16]{}\mathrel{=}\Varid{parseF}\;\Varid{xs}\mathrel{\hstretch{0.7}{>\!\!>\!\!=}}\Varid{stepM}\;\Varid{x}~~,{}\<[E]%
\\
\>[B]{}\hsindent{3}{}\<[3]%
\>[3]{}\mathbf{where}\;{}\<[10]%
\>[10]{}\Varid{stepM}\;\text{\ttfamily ')'}\;\Varid{ts}{}\<[34]%
\>[34]{}\mathrel{=}\Conid{Just}\;(\Conid{Nul}\mathbin{:}\Varid{ts}){}\<[E]%
\\
\>[10]{}\Varid{stepM}\;\text{\ttfamily '('}\;[\mskip1.5mu \Varid{t}\mskip1.5mu]{}\<[34]%
\>[34]{}\mathrel{=}\Conid{Nothing}{}\<[E]%
\\
\>[10]{}\Varid{stepM}\;\text{\ttfamily '('}\;(\Varid{t}\mathbin{:}\Varid{u}\mathbin{:}\Varid{ts}){}\<[34]%
\>[34]{}\mathrel{=}\Conid{Just}\;(\Conid{Bin}\;\Varid{t}\;\Varid{u}\mathbin{:}\Varid{ts})~~,{}\<[E]%
\ColumnHook
\end{hscode}\resethooks
where \ensuremath{\Varid{stepM}} is monadified \ensuremath{\Varid{step}} --- for the case \ensuremath{[\mskip1.5mu \Varid{t}\mskip1.5mu]} missing in \ensuremath{\Varid{step}} we return \ensuremath{\Conid{Nothing}}.

To relate \ensuremath{\Varid{parseF}} to \ensuremath{\Varid{parse}}, notice that \ensuremath{\Varid{prF}\;[\mskip1.5mu \Varid{t}\mskip1.5mu]\mathrel{=}\Varid{pr}\;\Varid{t}}.
We therefore have
\begin{hscode}\SaveRestoreHook
\column{B}{@{}>{\hspre}l<{\hspost}@{}}%
\column{14}{@{}>{\hspre}l<{\hspost}@{}}%
\column{E}{@{}>{\hspre}l<{\hspost}@{}}%
\>[B]{}\Varid{parse}\mathbin{::}\Conid{String}\to \Conid{Maybe}\;\Conid{Tree}{}\<[E]%
\\
\>[B]{}\Varid{parse}\mathrel{=}\Varid{unwrapM}\mathrel{\hstretch{0.7}{<\!\!\!=\!\!<}}\Varid{parseF}~~,{}\<[E]%
\\[\blanklineskip]%
\>[B]{}\Varid{unwrapM}\;[\mskip1.5mu \Varid{t}\mskip1.5mu]{}\<[14]%
\>[14]{}\mathrel{=}\Conid{Just}\;\Varid{t}{}\<[E]%
\\
\>[B]{}\Varid{unwrapM}\;\anonymous {}\<[14]%
\>[14]{}\mathrel{=}\Conid{Nothing}~~.{}\<[E]%
\ColumnHook
\end{hscode}\resethooks
where \ensuremath{(\mathrel{\hstretch{0.7}{<\!\!\!=\!\!<}})\mathbin{::}(\Varid{b}\to \Conid{M}\;\Varid{c})\to (\Varid{a}\to \Conid{M}\;\Varid{b})\to (\Varid{a}\to \Conid{M}\;\Varid{c})} is (reversed) Kleisli composition.
That is, \ensuremath{\Varid{parse}} calls \ensuremath{\Varid{parseF}}, and declares success only when the input can be parsed into a single tree.

\section{Longest Balanced Prefix in a Fold}
\label{sec:foldify}

Recall our objective at the close of Section \ref{sec:inits-tails}:
to compute
\ensuremath{\Varid{lbp}\mathrel{=}\Varid{maxBy}\;\Varid{size}\mathbin{\cdot}\Varid{filtJust}\mathbin{\cdot}\Varid{map}\;\Varid{parse}\mathbin{\cdot}\Varid{inits}} in a right fold,
in order to obtain a faster algorithm using the scan lemma.
Now that we have \ensuremath{\Varid{parse}\mathrel{=}\Varid{unwrapM}\mathrel{\hstretch{0.7}{<\!\!\!=\!\!<}}\Varid{parseF}} where \ensuremath{\Varid{parseF}} is a right fold,
we perform some initial calculation whose
purpose is to factor the postprocessing \ensuremath{\Varid{unwrapM}} out of the main computation:
\begin{hscode}\SaveRestoreHook
\column{B}{@{}>{\hspre}l<{\hspost}@{}}%
\column{7}{@{}>{\hspre}l<{\hspost}@{}}%
\column{9}{@{}>{\hspre}l<{\hspost}@{}}%
\column{E}{@{}>{\hspre}l<{\hspost}@{}}%
\>[7]{}\Varid{maxBy}\;\Varid{size}\mathbin{\cdot}\Varid{filtJust}\mathbin{\cdot}\Varid{map}\;\Varid{parse}\mathbin{\cdot}\Varid{inits}{}\<[E]%
\\
\>[B]{}\mathbin{=}{}\<[9]%
\>[9]{}\mbox{\commentbegin  since \ensuremath{\Varid{parse}\mathrel{=}\Varid{unwrapM}\mathrel{\hstretch{0.7}{<\!\!\!=\!\!<}}\Varid{parseF}}  \commentend}{}\<[E]%
\\
\>[B]{}\hsindent{7}{}\<[7]%
\>[7]{}\Varid{maxBy}\;\Varid{size}\mathbin{\cdot}\Varid{filtJust}\mathbin{\cdot}\Varid{map}\;(\Varid{unwrapM}\mathrel{\hstretch{0.7}{<\!\!\!=\!\!<}}\Varid{parseF})\mathbin{\cdot}\Varid{inits}{}\<[E]%
\\
\>[B]{}\mathbin{=}{}\<[9]%
\>[9]{}\mbox{\commentbegin  since \ensuremath{(\Varid{f}\mathrel{\hstretch{0.7}{<\!\!\!=\!\!<}}\Varid{g})\;\Varid{x}\mathrel{=}\Varid{f}\mathbin{\hstretch{0.7}{=\!\!<\!\!<}}\Varid{g}\;\Varid{x}}, map fusion (backwards)  \commentend}{}\<[E]%
\\
\>[B]{}\hsindent{7}{}\<[7]%
\>[7]{}\Varid{maxBy}\;\Varid{size}\mathbin{\cdot}\Varid{filtJust}\mathbin{\cdot}\Varid{map}\;(\Varid{unwrapM}\mathbin{\hstretch{0.7}{=\!\!<\!\!<}})\mathbin{\cdot}\Varid{map}\;\Varid{parseF}\mathbin{\cdot}\Varid{inits}{}\<[E]%
\\
\>[B]{}\mathbin{=}{}\<[9]%
\>[9]{}\mbox{\commentbegin  since \ensuremath{\Varid{filtJust}\mathbin{\cdot}\Varid{map}\;(\Varid{unwrapM}\mathbin{\hstretch{0.7}{=\!\!<\!\!<}})\mathrel{=}\Varid{map}\;\Varid{unwrap}\mathbin{\cdot}\Varid{filtJust}}, see below  \commentend}{}\<[E]%
\\
\>[B]{}\hsindent{7}{}\<[7]%
\>[7]{}\Varid{maxBy}\;\Varid{size}\mathbin{\cdot}\Varid{map}\;\Varid{unwrap}\mathbin{\cdot}\Varid{filtJust}\mathbin{\cdot}\Varid{map}\;\Varid{parseF}\mathbin{\cdot}\Varid{inits}{}\<[E]%
\\
\>[B]{}\mathbin{=}{}\<[9]%
\>[9]{}\mbox{\commentbegin  since \ensuremath{\Varid{maxBy}\;\Varid{f}\mathbin{\cdot}\Varid{map}\;\Varid{g}\mathrel{=}\Varid{g}\mathbin{\cdot}\Varid{maxBy}\;(\Varid{f}\mathbin{\cdot}\Varid{g})}  \commentend}{}\<[E]%
\\
\>[B]{}\hsindent{7}{}\<[7]%
\>[7]{}\Varid{unwrap}\mathbin{\cdot}\Varid{maxBy}\;(\Varid{size}\mathbin{\cdot}\Varid{unwrap})\mathbin{\cdot}\Varid{filtJust}\mathbin{\cdot}\Varid{map}\;\Varid{parseF}\mathbin{\cdot}\Varid{inits}~~.{}\<[E]%
\ColumnHook
\end{hscode}\resethooks
In the penultimate step \ensuremath{(\Varid{unwrapM}\mathbin{\hstretch{0.7}{=\!\!<\!\!<}})} is moved leftwards past \ensuremath{\Varid{filtJust}} and becomes \ensuremath{\Varid{unwrap}\mathbin{::}\Conid{Forest}\to \Conid{Tree}}, defined by:
\begin{hscode}\SaveRestoreHook
\column{B}{@{}>{\hspre}l<{\hspost}@{}}%
\column{13}{@{}>{\hspre}l<{\hspost}@{}}%
\column{E}{@{}>{\hspre}l<{\hspost}@{}}%
\>[B]{}\Varid{unwrap}\;[\mskip1.5mu \Varid{t}\mskip1.5mu]{}\<[13]%
\>[13]{}\mathrel{=}\Varid{t}{}\<[E]%
\\
\>[B]{}\Varid{unwrap}\;\anonymous {}\<[13]%
\>[13]{}\mathrel{=}\Conid{Nul}~~.{}\<[E]%
\ColumnHook
\end{hscode}\resethooks

Recall that \ensuremath{\Varid{inits}\mathrel{=}\Varid{foldr}\;(\lambda \Varid{x}\;\Varid{xss}\to [\mskip1.5mu \mskip1.5mu]\mathbin{:}\Varid{map}\;(\Varid{x}\mathbin{:})\;\Varid{xss})\;[\mskip1.5mu [\mskip1.5mu \mskip1.5mu]\mskip1.5mu]}.
The aim now is to fuse \ensuremath{\Varid{map}\;\Varid{parseF}}, \ensuremath{\Varid{filtJust}}, and \ensuremath{\Varid{maxBy}\;(\Varid{size}\mathbin{\cdot}\Varid{unwrap})} with \ensuremath{\Varid{inits}}.

By Theorem~\ref{thm:foldr-fusion}, to fuse \ensuremath{\Varid{map}\;\Varid{parseF}} with \ensuremath{\Varid{inits}}, we need to construct \ensuremath{\Varid{g}} that meets the fusion condition:
\begin{hscode}\SaveRestoreHook
\column{B}{@{}>{\hspre}l<{\hspost}@{}}%
\column{E}{@{}>{\hspre}l<{\hspost}@{}}%
\>[B]{}\Varid{map}\;\Varid{parseF}\;([\mskip1.5mu \mskip1.5mu]\mathbin{:}\Varid{map}\;(\Varid{x}\mathbin{:})\;\Varid{xss})\mathrel{=}\Varid{g}\;\Varid{x}\;(\Varid{map}\;\Varid{parseF}\;\Varid{xss})~~.{}\<[E]%
\ColumnHook
\end{hscode}\resethooks
Now that we know that \ensuremath{\Varid{parseF}} is a fold, the calculation goes:
\begin{hscode}\SaveRestoreHook
\column{B}{@{}>{\hspre}l<{\hspost}@{}}%
\column{7}{@{}>{\hspre}l<{\hspost}@{}}%
\column{8}{@{}>{\hspre}l<{\hspost}@{}}%
\column{E}{@{}>{\hspre}l<{\hspost}@{}}%
\>[7]{}\Varid{map}\;\Varid{parseF}\;([\mskip1.5mu \mskip1.5mu]\mathbin{:}\Varid{map}\;(\Varid{x}\mathbin{:})\;\Varid{xss}){}\<[E]%
\\
\>[B]{}\mathbin{=}{}\<[8]%
\>[8]{}\mbox{\commentbegin  definitions of \ensuremath{\Varid{map}} and \ensuremath{\Varid{parseF}}  \commentend}{}\<[E]%
\\
\>[B]{}\hsindent{7}{}\<[7]%
\>[7]{}\Conid{Just}\;[\mskip1.5mu \Conid{Nul}\mskip1.5mu]\mathbin{:}\Varid{map}\;(\Varid{parseF}\mathbin{\cdot}(\Varid{x}\mathbin{:}))\;\Varid{xss}{}\<[E]%
\\
\>[B]{}\mathbin{=}{}\<[8]%
\>[8]{}\mbox{\commentbegin  the \ensuremath{\Varid{foldr}} definition of \ensuremath{\Varid{parseF}}  \commentend}{}\<[E]%
\\
\>[B]{}\hsindent{7}{}\<[7]%
\>[7]{}\Conid{Just}\;[\mskip1.5mu \Conid{Nul}\mskip1.5mu]\mathbin{:}\Varid{map}\;(\lambda \Varid{ts}\to \Varid{parseF}\;\Varid{ts}\mathrel{\hstretch{0.7}{>\!\!>\!\!=}}\Varid{stepM}\;\Varid{x})\;\Varid{xss}{}\<[E]%
\\
\>[B]{}\mathbin{=}{}\<[8]%
\>[8]{}\mbox{\commentbegin  \ensuremath{\Varid{map}}-fusion (backwards)  \commentend}{}\<[E]%
\\
\>[B]{}\hsindent{7}{}\<[7]%
\>[7]{}\Conid{Just}\;[\mskip1.5mu \Conid{Nul}\mskip1.5mu]\mathbin{:}\Varid{map}\;(\mathrel{\hstretch{0.7}{>\!\!>\!\!=}}\Varid{stepM}\;\Varid{x})\;(\Varid{map}\;\Varid{parseF}\;\Varid{xss})~~.{}\<[E]%
\ColumnHook
\end{hscode}\resethooks
Therefore we have
\begin{hscode}\SaveRestoreHook
\column{B}{@{}>{\hspre}l<{\hspost}@{}}%
\column{4}{@{}>{\hspre}l<{\hspost}@{}}%
\column{E}{@{}>{\hspre}l<{\hspost}@{}}%
\>[B]{}\Varid{map}~\Varid{parseF}\mathbin{\cdot}\Varid{inits}\mathbin{::}\Conid{String}\to [\mskip1.5mu \Conid{Maybe}\;\Conid{Forest}\mskip1.5mu]{}\<[E]%
\\
\>[B]{}\Varid{map}~\Varid{parseF}\mathbin{\cdot}\Varid{inits}\mathrel{=}{}\<[E]%
\\
\>[B]{}\hsindent{4}{}\<[4]%
\>[4]{}\Varid{foldr}\;(\lambda \Varid{x}\;\Varid{tss}\to \Conid{Just}\;[\mskip1.5mu \Conid{Nul}\mskip1.5mu]\mathbin{:}\Varid{map}\;(\mathrel{\hstretch{0.7}{>\!\!>\!\!=}}\Varid{stepM}\;\Varid{x})\;\Varid{tss})\;[\mskip1.5mu \Conid{Just}\;[\mskip1.5mu \Conid{Nul}\mskip1.5mu]\mskip1.5mu]~~.{}\<[E]%
\ColumnHook
\end{hscode}\resethooks

Next, we fuse \ensuremath{\Varid{filtJust}} with \ensuremath{\Varid{map}\;\Varid{parseF}\mathbin{\cdot}\Varid{inits}} by Theorem~\ref{thm:foldr-fusion}.
After some calculations, we get:
\begin{hscode}\SaveRestoreHook
\column{B}{@{}>{\hspre}l<{\hspost}@{}}%
\column{5}{@{}>{\hspre}l<{\hspost}@{}}%
\column{12}{@{}>{\hspre}l<{\hspost}@{}}%
\column{E}{@{}>{\hspre}l<{\hspost}@{}}%
\>[B]{}\Varid{filtJust}\mathbin{\cdot}\Varid{map}~\Varid{parseF}\mathbin{\cdot}\Varid{inits}\mathbin{::}\Conid{String}\to [\mskip1.5mu \Conid{Forest}\mskip1.5mu]{}\<[E]%
\\
\>[B]{}\Varid{filtJust}\mathbin{\cdot}\Varid{map}~\Varid{parseF}\mathbin{\cdot}\Varid{inits}\mathrel{=}\Varid{foldr}\;(\lambda \Varid{x}\;\Varid{tss}\to [\mskip1.5mu \Conid{Nul}\mskip1.5mu]\mathbin{:}\Varid{extend}\;\Varid{x}\;\Varid{tss})\;[\mskip1.5mu [\mskip1.5mu \Conid{Nul}\mskip1.5mu]\mskip1.5mu]~~,{}\<[E]%
\\
\>[B]{}\hsindent{5}{}\<[5]%
\>[5]{}\mathbf{where}\;{}\<[12]%
\>[12]{}\Varid{extend}\;\text{\ttfamily ')'}\;\Varid{tts}\mathrel{=}\Varid{map}\;(\Conid{Nul}\mathbin{:})\;\Varid{tts}{}\<[E]%
\\
\>[12]{}\Varid{extend}\;\text{\ttfamily '('}\;\Varid{tts}\mathrel{=}[\mskip1.5mu (\Conid{Bin}\;\Varid{t}\;\Varid{u}\mathbin{:}\Varid{ts})\mid (\Varid{t}\mathbin{:}\Varid{u}\mathbin{:}\Varid{ts})\leftarrow \Varid{tts}\mskip1.5mu]~~.{}\<[E]%
\ColumnHook
\end{hscode}\resethooks
After the fusion we need not keep the \ensuremath{\Conid{Nothing}} entries in the fold;
the computation returns a collection of forests.
If the next character is \ensuremath{\text{\ttfamily ')'}}, we append \ensuremath{\Conid{Nul}} to every forest.
If the next entry is \ensuremath{\text{\ttfamily '('}}, we choose those forests having at least two trees, and combine them ---
the list comprehension keeps only the forests that match the pattern \ensuremath{(\Varid{t}\mathbin{:}\Varid{u}\mathbin{:}\Varid{ts})} and throws away those do not.
Note that \ensuremath{[\mskip1.5mu \Conid{Nul}\mskip1.5mu]}, to which the empty string is parsed, is always added to the collection of forests.

\begin{figure}[t]
{\small
\begin{center}
\begin{tabular}{lll}
\ensuremath{\Varid{inits}}    & \ensuremath{\Varid{map}\;\Varid{parseF}}         & \ensuremath{\Varid{filtJust}} \\
\hline
\ensuremath{\text{\ttfamily \char34 \char34}}       & \ensuremath{\Conid{J}\;[\mskip1.5mu \Conid{N}\mskip1.5mu]}              & \ensuremath{[\mskip1.5mu \Conid{N}\mskip1.5mu]} \\
\ensuremath{\text{\ttfamily \char34 (\char34}}      & \ensuremath{\Conid{Nothing}}            &  \\
\ensuremath{\text{\ttfamily \char34 ()\char34}}     & \ensuremath{\Conid{J}\;[\mskip1.5mu \Conid{B}\;\Conid{N}\;\Conid{N}\mskip1.5mu]}          & \ensuremath{[\mskip1.5mu \Conid{B}\;\Conid{N}\;\Conid{N}\mskip1.5mu]} \\
\ensuremath{\text{\ttfamily \char34 ())\char34}}    & \ensuremath{\Conid{J}\;[\mskip1.5mu \Conid{B}\;\Conid{N}\;\Conid{N},\Conid{N}\mskip1.5mu]}        & \ensuremath{[\mskip1.5mu \Conid{B}\;\Conid{N}\;\Conid{N},\Conid{N}\mskip1.5mu]} \\
\ensuremath{\text{\ttfamily \char34 ())(\char34}}   & \ensuremath{\Conid{Nothing}}            &  \\
\ensuremath{\text{\ttfamily \char34 ())()\char34}}  & \ensuremath{\Conid{J}\;[\mskip1.5mu \Conid{B}\;\Conid{N}\;\Conid{N},\Conid{B}\;\Conid{N}\;\Conid{N}\mskip1.5mu]}    & \ensuremath{[\mskip1.5mu \Conid{B}\;\Conid{N}\;\Conid{N},\Conid{B}\;\Conid{N}\;\Conid{N}\mskip1.5mu]}\\
\ensuremath{\text{\ttfamily \char34 ())()(\char34}} & \ensuremath{\Conid{Nothing}}            &
\end{tabular}
\end{center}
}
\caption{Results of \ensuremath{\Varid{parseF}} and \ensuremath{\Varid{filtJust}} for prefixes of \ensuremath{\text{\ttfamily \char34 ())()(\char34}}.}
\label{fig:FiltParseFExample}
\end{figure}

To think about how to deal with \ensuremath{\Varid{unwrap}\mathbin{\cdot}\Varid{maxBy}\;(\Varid{size}\mathbin{\cdot}\Varid{unwrap})}, we consider an example.
Figure~\ref{fig:FiltParseFExample} shows the results of \ensuremath{\Varid{map}\;\Varid{parseF}} and \ensuremath{\Varid{filtJust}} for prefixes of \ensuremath{\text{\ttfamily \char34 ())()(\char34}},
where \ensuremath{\Conid{Just}}, \ensuremath{\Conid{Nul}}, and \ensuremath{\Conid{Bin}} are respectively abbreviated to \ensuremath{\Conid{J}}, \ensuremath{\Conid{N}}, and \ensuremath{\Conid{B}}.
The function \ensuremath{\Varid{maxBy}\;(\Varid{size}\mathbin{\cdot}\Varid{unwrap})} chooses between \ensuremath{[\mskip1.5mu \Conid{N}\mskip1.5mu]} and \ensuremath{[\mskip1.5mu \Conid{B}\;\Conid{N}\;\Conid{N}\mskip1.5mu]}, the two parses resulting in single trees, and returns \ensuremath{[\mskip1.5mu \Conid{B}\;\Conid{N}\;\Conid{N}\mskip1.5mu]}.
However, notice that \ensuremath{\Conid{B}\;\Conid{N}\;\Conid{N}} is also the head of \ensuremath{[\mskip1.5mu \Conid{B}\;\Conid{N}\;\Conid{N},\Conid{B}\;\Conid{N}\;\Conid{N}\mskip1.5mu]}, the last forest returned by \ensuremath{\Varid{filtJust}}.
In general, the largest singleton parse tree will also present in the head of the last forest returned by \ensuremath{\Varid{filtJust}\mathbin{\cdot}\Varid{map}\;\Varid{parseF}\mathbin{\cdot}\Varid{inits}}.
One can intuitively see why: if we print them both, the former is a prefix of the latter.
Therefore, \ensuremath{\Varid{unwrap}\mathbin{\cdot}\Varid{maxBy}\;(\Varid{size}\mathbin{\cdot}\Varid{unwrap})} can be replaced by \ensuremath{\Varid{head}\mathbin{\cdot}\Varid{last}}.

To fuse \ensuremath{\Varid{last}} with \ensuremath{\Varid{filtJust}\mathbin{\cdot}\Varid{map}\;\Varid{parseF}\mathbin{\cdot}\Varid{inits}} by Theorem~\ref{thm:foldr-fusion}, we need to construct a function \ensuremath{\Varid{step}} that satisfies the fusion condition
\begin{hscode}\SaveRestoreHook
\column{B}{@{}>{\hspre}l<{\hspost}@{}}%
\column{E}{@{}>{\hspre}l<{\hspost}@{}}%
\>[B]{}\Varid{last}\;([\mskip1.5mu \Conid{Nul}\mskip1.5mu]\mathbin{:}\Varid{extend}\;\Varid{x}\;\Varid{tss})\mathrel{=}\Varid{step}\;\Varid{x}\;(\Varid{last}\;\Varid{tss})~~,{}\<[E]%
\ColumnHook
\end{hscode}\resethooks
where \ensuremath{\Varid{tss}} is a non-empty list of forests.
The case when \ensuremath{\Varid{x}\mathrel{=}\text{\ttfamily ')'}} is easy --- choosing \ensuremath{\Varid{step}\;\text{\ttfamily ')'}\;\Varid{ts}\mathrel{=}\Conid{Nul}\mathbin{:}\Varid{ts}} will do the job.
For the case when \ensuremath{\Varid{x}\mathrel{=}\text{\ttfamily '('}} we need to analyse the result of \ensuremath{\Varid{last}\;\Varid{tss}},
and use the property that forests in \ensuremath{\Varid{tss}} are ordered in ascending lengths.
\begin{compactenum}[a)]
\item If \ensuremath{\Varid{last}\;\Varid{tss}\mathrel{=}[\mskip1.5mu \Varid{t}\mskip1.5mu]}, a forest having only one tree, there are no forest in \ensuremath{\Varid{tss}} that contains two or more trees. Therefore \ensuremath{\Varid{extend}\;\text{\ttfamily '('}\;\Varid{tss}} returns an empty list, and \ensuremath{\Varid{last}\;([\mskip1.5mu \Conid{Nul}\mskip1.5mu]\mathbin{:}\Varid{extend}\;\text{\ttfamily '('}\;\Varid{tss})\mathrel{=}[\mskip1.5mu \Conid{Nul}\mskip1.5mu]}.
\item Otherwise, \ensuremath{\Varid{extend}\;\text{\ttfamily '('}\;\Varid{tss}} would not be empty, and \ensuremath{\Varid{last}\;([\mskip1.5mu \Conid{Nul}\mskip1.5mu]\mathbin{:}\Varid{extend}\;\Varid{x}\;\Varid{tss})\mathrel{=}\Varid{last}\;(\Varid{extend}\;\Varid{x}\;\Varid{tss})}. We may then combine the first two trees, as \ensuremath{\Varid{extend}} would do.
\end{compactenum}
In summary, we have
\begin{hscode}\SaveRestoreHook
\column{B}{@{}>{\hspre}l<{\hspost}@{}}%
\column{3}{@{}>{\hspre}l<{\hspost}@{}}%
\column{10}{@{}>{\hspre}l<{\hspost}@{}}%
\column{29}{@{}>{\hspre}l<{\hspost}@{}}%
\column{E}{@{}>{\hspre}l<{\hspost}@{}}%
\>[B]{}\Varid{last}\mathbin{\cdot}\Varid{filtJust}\mathbin{\cdot}\Varid{map}~\Varid{parseF}\mathbin{\cdot}\Varid{inits}\mathbin{::}\Conid{String}\to \Conid{Forest}{}\<[E]%
\\
\>[B]{}\Varid{last}\mathbin{\cdot}\Varid{filtJust}\mathbin{\cdot}\Varid{map}~\Varid{parseF}\mathbin{\cdot}\Varid{inits}\mathrel{=}\Varid{foldr}\;\Varid{step}\;[\mskip1.5mu \Conid{Nul}\mskip1.5mu]~~,{}\<[E]%
\\
\>[B]{}\hsindent{3}{}\<[3]%
\>[3]{}\mathbf{where}\;{}\<[10]%
\>[10]{}\Varid{step}\;\text{\ttfamily ')'}\;\Varid{ts}{}\<[29]%
\>[29]{}\mathrel{=}\Conid{Nul}\mathbin{:}\Varid{ts}{}\<[E]%
\\
\>[10]{}\Varid{step}\;\text{\ttfamily '('}\;[\mskip1.5mu \Varid{t}\mskip1.5mu]{}\<[29]%
\>[29]{}\mathrel{=}[\mskip1.5mu \Conid{Nul}\mskip1.5mu]{}\<[E]%
\\
\>[10]{}\Varid{step}\;\text{\ttfamily '('}\;(\Varid{t}\mathbin{:}\Varid{u}\mathbin{:}\Varid{ts}){}\<[29]%
\>[29]{}\mathrel{=}\Conid{Bin}\;\Varid{t}\;\Varid{u}\mathbin{:}\Varid{ts}~~,{}\<[E]%
\ColumnHook
\end{hscode}\resethooks
which is now a total function on strings of parentheses.

The function derived above turns out to be \ensuremath{{\Varid{prF}}^{\hstretch{0.5}{-}1}} with one additional case (\ensuremath{\Varid{step}\;\text{\ttfamily '('}\;[\mskip1.5mu \Varid{t}\mskip1.5mu]\mathrel{=}[\mskip1.5mu \Conid{Nul}\mskip1.5mu]}). What we have done in this section can be seen as justifying this extra case (which is a result of case (1) in the fusion of \ensuremath{\Varid{last}}), which is not as trivial as one might think.

\section{Wrapping Up}
\label{sec:wrap}

We can finally resume the main derivation in Section~\ref{sec:inits-tails}:
\begin{hscode}\SaveRestoreHook
\column{B}{@{}>{\hspre}l<{\hspost}@{}}%
\column{7}{@{}>{\hspre}l<{\hspost}@{}}%
\column{9}{@{}>{\hspre}l<{\hspost}@{}}%
\column{E}{@{}>{\hspre}l<{\hspost}@{}}%
\>[7]{}\Varid{maxBy}\;\Varid{size}\mathbin{\cdot}\Varid{map}\;(\Varid{maxBy}\;\Varid{size}\mathbin{\cdot}\Varid{filtJust}\mathbin{\cdot}\Varid{map}\;\Varid{parse}\mathbin{\cdot}\Varid{inits})\mathbin{\cdot}\Varid{tails}{}\<[E]%
\\
\>[B]{}\mathbin{=}{}\<[9]%
\>[9]{}\mbox{\commentbegin  Section~\ref{sec:foldify}: \ensuremath{\Varid{lbp}\mathrel{=}\Varid{head}\mathbin{\cdot}\Varid{foldr}\;\Varid{step}\;[\mskip1.5mu \Conid{Nul}\mskip1.5mu]}  \commentend}{}\<[E]%
\\
\>[B]{}\hsindent{7}{}\<[7]%
\>[7]{}\Varid{maxBy}\;\Varid{size}\mathbin{\cdot}\Varid{map}\;(\Varid{head}\mathbin{\cdot}\Varid{foldr}\;\Varid{step}\;[\mskip1.5mu \Conid{Nul}\mskip1.5mu])\mathbin{\cdot}\Varid{tails}{}\<[E]%
\\
\>[B]{}\mathbin{=}{}\<[9]%
\>[9]{}\mbox{\commentbegin  \ensuremath{\Varid{map}}-fusion reversed, Lemma~\ref{lma:scan-lemma}  \commentend}{}\<[E]%
\\
\>[B]{}\hsindent{7}{}\<[7]%
\>[7]{}\Varid{maxBy}\;\Varid{size}\mathbin{\cdot}\Varid{map}\;\Varid{head}\mathbin{\cdot}\Varid{scanr}\;\Varid{step}\;[\mskip1.5mu \Conid{Nul}\mskip1.5mu]~~.{}\<[E]%
\ColumnHook
\end{hscode}\resethooks

We have therefore derived:
\begin{hscode}\SaveRestoreHook
\column{B}{@{}>{\hspre}l<{\hspost}@{}}%
\column{E}{@{}>{\hspre}l<{\hspost}@{}}%
\>[B]{}\Varid{lbs}\mathbin{::}\Conid{String}\to \Conid{Tree}{}\<[E]%
\\
\>[B]{}\Varid{lbs}\mathrel{=}\Varid{maxBy}\;\Varid{size}\mathbin{\cdot}\Varid{map}\;\Varid{head}\mathbin{\cdot}\Varid{scanr}\;\Varid{step}\;[\mskip1.5mu \Conid{Nul}\mskip1.5mu]~~,{}\<[E]%
\ColumnHook
\end{hscode}\resethooks
where \ensuremath{\Varid{step}} is as defined in the end of Section~\ref{sec:foldify}.
To avoid recomputing the sizes in \ensuremath{\Varid{maxBy}\;\Varid{size}}, we can annotate each tree by its size: letting \ensuremath{\Conid{Forest}\mathrel{=}[\mskip1.5mu (\Conid{Tree},\Conid{Int})\mskip1.5mu]}, resulting in an algorithm that runs in linear-time:
\begin{hscode}\SaveRestoreHook
\column{B}{@{}>{\hspre}l<{\hspost}@{}}%
\column{4}{@{}>{\hspre}l<{\hspost}@{}}%
\column{11}{@{}>{\hspre}l<{\hspost}@{}}%
\column{25}{@{}>{\hspre}l<{\hspost}@{}}%
\column{E}{@{}>{\hspre}l<{\hspost}@{}}%
\>[B]{}\Varid{lbs}\mathbin{::}\Conid{String}\to \Conid{Tree}{}\<[E]%
\\
\>[B]{}\Varid{lbs}\mathrel{=}\Varid{fst}\mathbin{\cdot}\Varid{maxBy}\;\Varid{snd}\mathbin{\cdot}\Varid{map}\;\Varid{head}\mathbin{\cdot}\Varid{scanr}\;\Varid{step}\;[\mskip1.5mu (\Conid{Nul},\mathrm{0})\mskip1.5mu]~~,{}\<[E]%
\\
\>[B]{}\hsindent{4}{}\<[4]%
\>[4]{}\mathbf{where}\;{}\<[11]%
\>[11]{}\Varid{step}\;\text{\ttfamily ')'}\;\Varid{ts}{}\<[25]%
\>[25]{}\mathrel{=}(\Conid{Nul},\mathrm{0})\mathbin{:}\Varid{ts}{}\<[E]%
\\
\>[11]{}\Varid{step}\;\text{\ttfamily '('}\;[\mskip1.5mu \Varid{t}\mskip1.5mu]{}\<[25]%
\>[25]{}\mathrel{=}[\mskip1.5mu (\Conid{Nul},\mathrm{0})\mskip1.5mu]{}\<[E]%
\\
\>[11]{}\Varid{step}\;\text{\ttfamily '('}\;((\Varid{t},\Varid{m})\mathbin{:}(\Varid{u},\Varid{n})\mathbin{:}\Varid{ts})\mathrel{=}(\Conid{Bin}\;\Varid{t}\;\Varid{u},\mathrm{2}\mathbin{+}\Varid{m}\mathbin{+}\Varid{n})\mathbin{:}\Varid{ts}~~.{}\<[E]%
\ColumnHook
\end{hscode}\resethooks
Finally, the size-only version can be obtained by fusing \ensuremath{\Varid{size}} into \ensuremath{\Varid{lbs}}.
It turns out that we do not need to keep the actual trees, but only their sizes ---
\ensuremath{\Conid{Forest}\mathrel{=}[\mskip1.5mu \Conid{Int}\mskip1.5mu]}:
\begin{hscode}\SaveRestoreHook
\column{B}{@{}>{\hspre}l<{\hspost}@{}}%
\column{3}{@{}>{\hspre}l<{\hspost}@{}}%
\column{10}{@{}>{\hspre}l<{\hspost}@{}}%
\column{24}{@{}>{\hspre}l<{\hspost}@{}}%
\column{E}{@{}>{\hspre}l<{\hspost}@{}}%
\>[B]{}\Varid{lbsl}\mathbin{::}\Conid{String}\to \Conid{Int}{}\<[E]%
\\
\>[B]{}\Varid{lbsl}\mathrel{=}\Varid{maximum}\mathbin{\cdot}\Varid{map}\;\Varid{head}\mathbin{\cdot}\Varid{scanr}\;\Varid{step}\;[\mskip1.5mu \mathrm{0}\mskip1.5mu]~~,{}\<[E]%
\\
\>[B]{}\hsindent{3}{}\<[3]%
\>[3]{}\mathbf{where}\;{}\<[10]%
\>[10]{}\Varid{step}\;\text{\ttfamily ')'}\;\Varid{ts}{}\<[24]%
\>[24]{}\mathrel{=}\mathrm{0}\mathbin{:}\Varid{ts}{}\<[E]%
\\
\>[10]{}\Varid{step}\;\text{\ttfamily '('}\;[\mskip1.5mu \Varid{t}\mskip1.5mu]{}\<[24]%
\>[24]{}\mathrel{=}[\mskip1.5mu \mathrm{0}\mskip1.5mu]{}\<[E]%
\\
\>[10]{}\Varid{step}\;\text{\ttfamily '('}\;(\Varid{m}\mathbin{:}\Varid{n}\mathbin{:}\Varid{ts})\mathrel{=}(\mathrm{2}\mathbin{+}\Varid{m}\mathbin{+}\Varid{n})\mathbin{:}\Varid{ts}~~.{}\<[E]%
\ColumnHook
\end{hscode}\resethooks

\begin{figure}
\begin{center}
\begin{tabular}{|r|rrrrrr|}
\hline
input size (M)      & 1 & 2 & 4 & 6 & 8 & 10 \\
\hline
user time (sec.)
& 0.52
& 1.25
& 2.38
& 3.20
& 4.74
& 5.50\\
\hline
\end{tabular}
\end{center}
\caption{Measured running time for some input sizes.}
\label{fig:experiments}
\end{figure}

We ran some simple experiments to measure the efficiency of the algorithm.
The test machine was a laptop computer with a Apple M1 chip (8 core, 3.2GHz) and 16GB RAM.
We ran \ensuremath{\Varid{lbs}} on randomly generated inputs containing 1, 2, 4, 6, 8, and 10 million  parentheses, and measured the user times.
The results, shown in Figure~\ref{fig:experiments}, confirmed the linear-time behaviour.

\section{Conclusions and Discussions}
\label{sec:conclude}

So we have derived a linear-time algorithm for solving the problem.
We find it an interesting journey because it relates two techniques:
prefix-suffix decomposition for solving segment problems, and the converse-of-a-function theorem for program inversion.

In Section~\ref{sec:partial-balance} we generalised from trees to forests.
Generalisations are common when applying the converse-of-a-function theorem.
It was observed that the trees in a forest are those along the left spine of the final tree, therefore such a generalisation is referred to as switching to a ``spine representation'' \citep{MuBird:03:Theory}.

What we derived in Section~\ref{sec:spine} and \ref{sec:foldify} is a compacted form of shift-reduce parsing, where the input is processed right-to-left.
The forest serves as the stack, but we do not need to push the parser state to the stack, as is done in shift-reduce parsing.
If we were to process the input in the more conventional left-to-right order, the corresponding grammar would be \ensuremath{\Conid{S}\to \epsilon\mid \Conid{S}\;(\Conid{S})}.
It is an SLR(1) grammar whose parse table contains 5 states.
Our program is much simpler.
A possible reason is that consecutive shifting and reducing are condensed into one step.
It is likely that parsing SLR(1) languages can be done in a fold.
The relationship between LR parsing and the converse-of-a-function theorem awaits further investigation.

There are certainly other ways to solve the problem.
For example, one may interpret a \ensuremath{\text{\ttfamily '('}} as a $-1$, and a \ensuremath{\text{\ttfamily ')'}} as a $+1$.
A left-partially balanced string would be a list whose right-to-left running sum is never negative.
One may then apply the method in \cite{Zantema:92:Longest} to find the longest such prefix for each suffix.
The result will be an algorihm that maintains the sum in a loop --- an approach that might be more commonly adopted by imperative programmers.
The problem can also be seen as an instance of \emph{maximum-marking problems} --- choosing elements in a data structure that meet a given criteria while maximising a cost function --- to which methods of \cite{SasanoHu:01:Generation} can be applied.

\paragraph*{Acknowledgements}~
The problem was suggested by Yi-Chia Chen.
The authors would like to thank our colleagues in the PLFM group in IIS, Academia Sinica, in particular Hsiang-Shang `Josh' Ko, Liang-Ting Chen, and Ting-Yan Lai, for valuable discussions.
Also thanks to Chung-Chieh Shan and Kim-Ee Yeoh for their advice on earlier drafts of this paper.
We are grateful to the reviewers of previous versions of this article, who gave detailed and constructive criticisms that helped a lot in improving this work.


\bibliographystyle{abbrvnat}

\appendix

\end{document}